\documentclass[aps,prl,twocolumn,showpacs]{revtex4}
\usepackage{graphics,graphicx}
\usepackage{color}
\usepackage{wrapfig}
\usepackage{enumerate}
\usepackage{amssymb,amsmath}
\usepackage{bm}

\def\w{\omega}

\def\sgn{{\rm sgn}}
\def\bk{{\bm k}}

\def\D2n{\Delta^2(s, \w_n)}
\def\enk{\varepsilon_{\bm k}}

\def\cks{c_{{\bm k}\sigma}}
\def\cdks{c^\dagger_{{\bm k}\sigma}}

\def\cdku{c^\dagger_{{\bm k}\uparrow}}
\def\cdkQu{c^\dagger_{{\bm k}+{\bm Q} \uparrow}}
\def\cdkd{c^\dagger_{{\bm k}\downarrow}}

\def\cku{c_{{\bm k}\uparrow}}
\def\ckd{c_{{\bm k}\downarrow}}
\def\ckQu{c_{{\bm k}+{\bm Q} \uparrow}}

\def\cdkQu{c^\dagger_{{\bm k}+{\bm Q} \uparrow}}

\def\cdmku{c^\dagger_{-{\bm k}\uparrow}}
\def\cdmkd{c^\dagger_{-{\bm k}\downarrow}}

\def\uk{u_{\bm k}}

\def\vk{v_{\bm k}}

\def\beq{\begin{equation}}
\def\seq{\end{equation}}
\def\beqs{\begin{eqnarray}}
\def\seqs{\end{eqnarray}}

\begin{document}
\unitlength = 1mm
\title{Antiferromagnetic Order in Pauli Limited Unconventional Superconductors}

\author{Yasuyuki~Kato$^1$,  C.~D.~Batista$^1$, I.~Vekhter$^2$}
\affiliation{$^1$Theoretical Division, Los Alamos National Laboratory, Los Alamos, NM 87545}
\affiliation{$^2$Department of Physics and Astronomy,
             Louisiana State University, Baton Rouge, Louisiana, 70803, USA}

\date{\today}
\pacs{71.27.+a, 74.20.Rp, 74.70.Tx}

\begin{abstract}
We develop a theory of the coexistence of superconductivity (SC) and antiferromagnetism (AFM) in  CeCoIn$_5$. We show that in Pauli-limited nodal superconductors the nesting of the quasi-particle pockets induced by Zeeman pair-breaking leads to  incommensurate AFM with the magnetic moment 
normal to the field. We compute the  phase diagram and find a first order transition to the normal state at low temperatures, absence of normal state AFM, and coexistence of SC and AFM at high fields, in agreement with experiments. We also predict
the existence of a new double-${\bm Q}$ magnetic phase.
\end{abstract}
\maketitle

The interplay between antiferromagnetism 
and unconventional superconductivity 
is one of the most intensely investigated topics in
correlated electron systems. The 115 family of heavy fermion compounds (CeMIn$_5$ where M=Co,Rh, Ir) exhibits many salient features of this interplay. CeCoIn$_5$ is a clean layered singlet $d$-wave superconductor whose upper critical field, $H_{c2}$, is determined by the Zeeman splitting of the electronic states (Pauli limiting), rather than by the orbital motion of the Cooper pairs.
This material
has an unusual low-temperature phase at fields just below $H_{c2}$, separated by a second order transition line from the low field superconductor, and by a first order transition from the normal (N) state  ($H\geq H_{c2}$)~\cite{ABianchi:2003a}. Initially this phase was conjectured~\cite{ABianchi:2003a} to be the first realization of the  Fulde-Ferrell-Larkin-Ovchinnikov (FFLO) state, where Cooper pairs acquire a finite center of mass momentum to counteract the pair-breaking due to Pauli limiting, and the superconducting order parameter oscillates in real space.

However, experiments present a more complex situation. The high-field N-SC
transition becomes first order at temperatures, $T$, higher than that of the onset of the new phase~\cite{ABianchi:2003a}. Specific heat and resistivity are anomalous just above $H_{c2}$, consistent with quantum critical behavior due to antiferromagnetic fluctuations~\cite{ABianchi:2003b,JPaglione:2003}, yet AFM was not found upon suppression of SC by several means~\cite{ABianchi:2003b,EDBauer:2005,FRonning:2006,CFMiclea:2006}. Finally, a series of NMR and neutron scattering measurements with ${\bm H}$ parallel to the layers established the existence of static AFM in the high-field 
phase~\cite{VMitrovic:2006,BLYoung:2007,MKenzelmann:2008,GKoutroulakis:2010,MKenzelmann:2010,EBlackburn:2010}. The ordered moment ${\bm m}\perp {\bm H}$
is modulated with an essentially field-independent wave-vector
${\bm Q}/(2\pi)=(q,q,0.5)$, $(q\simeq0.44 )$
\cite{MKenzelmann:2008,MKenzelmann:2010}. This unusual phase challenges our understanding of the connection between SC and AFM.

In this Letter we show that the near-perfect nesting of the quasiparticle pockets created by Zeeman pair-breaking  near the nodal regions creates the conditions for the antiferromagnetic instability even when the normal state Fermi surface is not nested.
This AFM instability is equivalent to equal spin pairing of Bogoliubov quasiparticles within each pocket. We compute the transition temperature both analytically in the weak-coupling limit (WCL) and numerically for a two-dimensional (2D) model. The resulting coexistence of AFM and SC appears in the low-temperature and high-field region of the phase diagram, while magnetic order does not appear in the normal state. Moreover, the  N-SC transition becomes first order at a temperature above that where the magnetic order first appears. These results are in agreement with the experimental observations on CeCoIn$_5$. The most favorable magnetic state
in the WCL is a double-${\bm Q}$ collinear structure at the incommensurate wave-vectors connecting opposite nodes, with  ${\bm m} \perp {\bm H}$, but a single-${\bm Q}$ state may be stabilized in a part of the phase diagram.

Previous theories focused on the field induced AFM and considered the single-${\bm Q}$ state only.  In some proposals the amplitude modulation of the superconducting order parameter, $\Delta(\bm r)$,  is essential and drives a spin density wave (SDW) order that varies on the same length scale~\cite{DFAgterberg:2009,YYanase:2009}. In other theories the SDW instability is not conditional on the oscillatory behavior of $\Delta(\bm r)$
~\cite{AAperis:2008,AAperis:2010,RIkeda:2010}.
In Refs.~\cite{AAperis:2008,AAperis:2010} the direction of the SDW magnetization was taken parallel to ${\bm H}$, and the phase diagram was obtained numerically for fixed (large) values of
interactions in different channels. We find that  the longitudinal antiferromagnetic susceptibility is gapped
under Zeeman field, but the transverse susceptibility is divergent. Hence AFM with ${\bm m} \perp {\bm H}$ appears
naturally. Landau expansion of the free energy in a model with critical magnetic fluctuations also supports ${\bm m} \perp {\bm H}$ for nodal superconductors~\cite{RIkeda:2010}. It was recently suggested~\cite{KSuzuki:2010} that an enhancement of the density of states above the normal state value due to combined effect of vortices and strong Pauli paramagnetism leads to the SDW instability, with no FFLO-like modulation of $\Delta(\bm r)$. 
In our theory it is the nesting between pockets, and therefore the joint density of states enhancement, that drives the magnetic transition. 
Finally, we propose the existence  of a double-${\bm Q}$ spin structure that differs from 
previous theories and
provides a  test of our results.

\begin{figure}[t]
\centerline{\includegraphics[angle=0,width=0.8\linewidth]{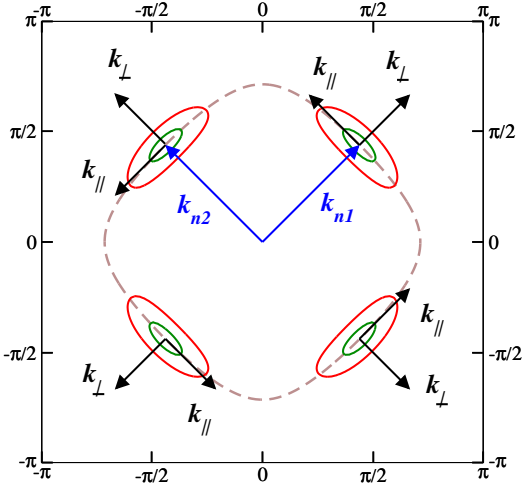}}
\caption{(Color online)
Fermi surface and the nodal pockets generated by Zeeman field for
the tight-binding model described in the text with $\Delta_0/t=0.7$, $h/t=0.3, 0.7$.
${\bm k}_{\parallel}$ and ${\bm k}_{\perp}$ correspond to the directions parallel and perpendicular
to the Fermi surface near each pocket.}
\label{fig:nodes}
\end{figure}
Since the main Fermi surface sheet of CeCoIn$_5$ is quasi-cylindrical, we consider a 
2D Hamiltonian
for a $d_{x^2-y^2}$ superconductor
with a Zeeman term  $h=g\mu_B H$,   
\begin{eqnarray}
    {\cal H}_{0} =\sum_{\bk \sigma} (\enk-h\sigma) \cdks \cks
    -\sum_{\bk}\left(\Delta_{\bm k}\cdku\cdmkd
    + {\rm H. c.}\right).
\nonumber
\end{eqnarray}
Here
$\varepsilon_{\bm k}$ is the band dispersion and $\Delta_{\bm k}$ is the $d_{x^2-y^2}$ order parameter.
We select the spin quantization axis along ${\bm H}$.
A Bogoliubov transformation $\gamma_{\bm k\uparrow}= \uk\cku-\vk^\star \cdmkd$,  $\gamma_{\bm k\downarrow}= \uk\ckd+\vk^\star \cdmku$ with
$2|\uk|^2 =\left[1+{\enk}/{E_{\bm k}}\right]$, $2|\vk|^2=\left[1-{\enk}/{E_{\bm k}}\right]$, and $E_{\bm k}=\sqrt{\enk^2+\Delta_{\bm k}^2}$  brings the Hamiltonian to the diagonal form
\begin{equation}
   \widetilde{\cal H}_0=\sum_{\bm k} \left\{\left[E_{\bm k}-h\right] \gamma^\dagger_{\bm k \uparrow}\gamma_{\bm k\uparrow}
  +\left[E_{\bm k}+ h\right]
    \gamma^\dagger_{-\bm k\downarrow}\gamma_{-\bm
    k\downarrow}\right\}\,.
    \label{eq:H0}
\end{equation}
The first term describes the creation of spin-polarized pockets of Bogoliubov quasiparticles by magnetic field. These pockets are unstable to the formation of AFM, and we first elucidate this instability in the WCL.

At low fields we linearize the dispersion in the near-nodal regions by taking $\enk\approx v_F k_\perp$ and $\Delta_{\bm k} \approx v_\Delta k_\|$ with $k_\perp$ ($k_\|$) measured from the node normal (parallel) to the Fermi surface (see Fig.~\ref{fig:nodes}) and obtain elliptical Fermi pockets $\sqrt{v_F^2 k_\perp^2+v_\Delta^2 k_\|^2}=h$. For a centrosymmetric system with pairs of nodes at momenta $\pm{\bm k}_{ni}$, the
low field pockets are perfectly nested at ${\bm Q}_i =2 {\bm k}_{ni}$ and ${\bar {\bm Q}}_i =-2 \bm k_{ni}$ ($i=1,2$, see Fig.\ref{fig:nodes}), even though the Fermi surface in the normal state is not. This nesting drives the instabilities of the superconducting state.

As 115 materials exhibit strong antiferromagnetic fluctuations, we assume that there is a residual interaction in this channel.
We start by considering the simplest AFM
with $\bm m\perp \bm H$: a single-${\bm Q}$ spiral
$ {\bm m}({\bm r}) =m_{{\bm Q}} (\cos\bm{Q \cdot r}, \sin\bm{Q \cdot r}, 0)$,
where ${\bm Q}\in\pm {\bm Q}_{i} $ is one of the four wave-vectors that are perfectly nested.
The mean field Hamiltonian
for this spiral order is
\begin{equation}
  {\cal H}_{AFM}=-J\sum_{\bm k} \left[m_{{\bm Q}} \cdkd c^{\;}_{{\bm k}+{\bm Q} \uparrow} +  m^*_{{\bm Q}} c^{\dagger}_{{\bm k}+{\bm Q}\uparrow} \ckd\right],
\label{eq:H_AFM}
\end{equation}
where $ m_{\bm Q}=\sum_{\boldsymbol k}\langle c^{\dagger}_{{\boldsymbol k} +{\boldsymbol Q} \uparrow } c^{\;}_{{\boldsymbol k} \downarrow} \rangle/N $, $J$ is the coupling constant, $\langle\ldots\rangle$ denotes a thermodynamic average, and $N$ is the number of lattices sites.

\begin{figure}[t]
\centerline{\includegraphics[width=0.9\linewidth]{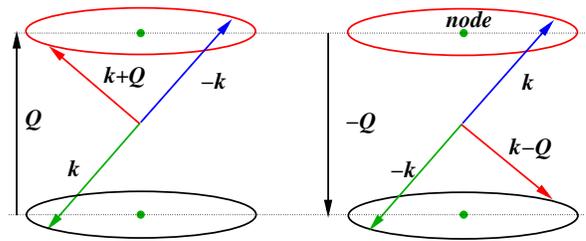}}
\caption{(Color online)
Conditions for pairing of the Fermions at the nodal pockets due to the term $c^\dagger_{{\bm k}+{\bm Q}_i \uparrow}\ckd$: a placeholder. Dotted line: Fermi surface, green dots: nodes. Left panel: If ${\bm k}$ is in the lower pocket, only the term with ${\bm Q}_i={\bm Q}$ satisfies the nesting condition and leads to pairing. Right panel: If ${\bm k}$ is in the upper pocket, only ${\bm Q}_i=-{\bm Q}$ is relevant.}
\label{fig:pairs}
\end{figure}

Since a Bogoliubov quasiparticle with a given spin at momentum $\bm k$
combines an electron state with the same spin at ${\bm k}$ with
a hole with opposite spin at $-{\bm k}$, the
AFM interaction in the particle-hole channel is manifested as a spin triplet superconducting instability of the Bogoliubov quasiparticles within each pocket. Transforming
Eq.~(\ref{eq:H_AFM}) into the basis of Bogoliubov states we find
\begin{eqnarray}
\nonumber
  \cdkQu\ckd =-u_{\bm k+\bm Q} v_{\bm k}^\star\gamma^\dagger_{\bm k+\bm Q \uparrow}
  \gamma^\dagger_{-\bm k\uparrow} + \mbox{irr.}\,.
  \label{eq:AFM2}
\end{eqnarray}
Note that we only need to consider the quasiparticles with the spin along the field (occupied states). The terms marked as irrelevant (irr.) include spin down Bogoliubov operators, and therefore do not have an expectation value since these excitations are gapped by $h$.  Consider the vicinity of the node $-{\bm Q}/2$. We have $\bm k=-{\bm Q}/2+\bm q$, so that $\bm k+\bm Q={\bm Q}/2+\bm q$, and we find
\begin{equation}
\nonumber
  \cdkQu\ckd=-u_{ \frac{\bm Q}{2}+{\bm q}} v_{-\frac{\bm Q}{2} +{\bm q}}^\star
  \gamma^\dagger_{\frac{\bm Q}{2}+{\bm q} \uparrow}\gamma^\dagger_{\frac{\bm Q}{2}
-{\bm q} \uparrow} + \mbox{irr.}\,,
\end{equation}
which describes the triplet pairing of ``Bogoliubons'' within the same pocket. Since $u_{\bm k+\bm Q} v_{\bm k}^\star\approx \sgn(k_\|)$ in the linearized approximation for a $d$-wave gap, we obtain
\begin{eqnarray}
  {\cal H}&=& \sum_{\bm q} \left[\left(E_{-\frac{\bm Q}{2}+\bm q}-h\right)\gamma^\dagger_{-\frac{\bm Q}{2}+\bm q \uparrow}\right.
  \gamma_{-\frac{\bm Q}{2}+\bm q\uparrow}
\nonumber
\\
& &\left. + J m^*_{\bm Q}  \gamma^\dagger_{\frac{\bm Q}{2}+\bm q \uparrow}\gamma^\dagger_{\frac{\bm Q}{2}-\bm q\uparrow} \sgn(q_\|)
+ {\rm H. c.}\right]\,,
\label{hap}
\end{eqnarray}
showing $p$-wave pairing symmetry. The sum in Eq.\eqref{hap} runs over wave-vectors ${\bm q}$ 
near the pocket
center, i.e. $|{\bm q}| \ll \pi/a$. This approximation is valid as long as ${\bm Q}$ is far
from the commensurate wave vector $(\pi,\pi)$. In that case, only the wave-vectors on the Fermi surface of the pocket
centered at $-{\bm Q}/2$ satisfy the nesting condition for the transferred momentum 
${\bm Q}$ (see Fig.\ref{fig:pairs}).

The critical temperature in the WCL is
\begin{equation}
  T_{\bm Q}\approx \sqrt{\omega_c h}\exp\left[-\frac{2\pi^2 v_F v_\Delta}{Jh}\right]\,,
  \label{eq:TQ}
\end{equation}
where $\omega_c$ is the high energy cutoff for the magnetic interaction $J$. This result is valid for $T_{\bm Q}\ll h\ll \{\omega_c,\Delta_0\}$,
and describes the instability of the uniform SC towards the AFM modulation.
It is likely that the exponentially small $\bm m$ at low $H$ is destroyed by the orbital effect (see below).
The order parameter that appears at $T_{\bm Q}$, 
$  \sum_{\bm q} \sgn(q_\|) \langle\gamma^\dagger_{{\bm Q}/2+\bm q \uparrow}\gamma^\dagger_{{\bm Q}/2-\bm q\uparrow}\rangle $,
has a staggered triplet superconducting component, $\Delta^t_{\bm Q} =
\sum_{\bm q} \delta^t_{\bm q} \langle c^\dagger_{-{\bm Q}/2+q \uparrow}c^\dagger_{-{\bm Q}/2-q \uparrow}\rangle $
with $\sgn(\delta^t_{\bm q})=\sgn(q_\|)$, in addition to the AFM
$m_{\bm Q}$ \cite{AAperis:2008,AAperis:2010}.
This triplet component is an inevitable companion of $\Delta_0$ and $m_{\bm Q}$ since the Ginzburg-Landau expansion allows for the invariants of the form ${\boldsymbol \Delta}_{\boldsymbol Q} \cdot  {\boldsymbol m}_{\boldsymbol {\bar Q}} \Delta^*_0 +  {\boldsymbol \Delta}^*_{\boldsymbol {\bar Q}} \cdot  {\boldsymbol m}_{\boldsymbol  Q} \Delta_0$, linear in ${\bm \Delta}_{\bm Q}$, where
${\bm \Delta}_{\bm Q}$ is the vector 
of three independent components of the
triplet superconducting order parameter ($2 \Delta^t_{\bm Q} = \Delta^x_{\bm Q} + i \Delta^y_{\bm Q} $). However, we expect $\Delta^t_{\bm Q}$ to be extraordinary fragile with respect to impurity scattering so it cannot be easily observed in CeCoIn$_5$.

\begin{figure}[t]
\centerline{\includegraphics[angle=270,trim =0 0 0 0,width=8cm,clip]{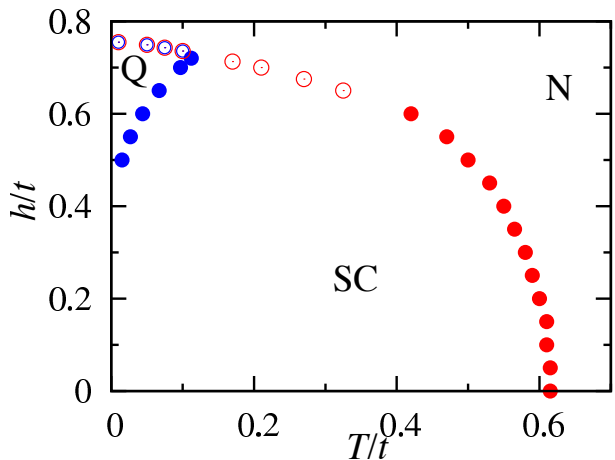}}
\caption{(Color online)
Phase diagram obtained for $V=3t$ and $J=3.5t$. Open and filled circles indicate first and second order phase transitions respectively.
The red circles are the boundary between the SC and N states, while the blue circles are the boundary for the magnetically
ordered phase.}
\label{fig:PD}
\end{figure}

To
go beyond WCL, we consider a mean field Hamiltonian
with both SC and AFM
interactions,
\begin{eqnarray}
  {\cal H}&=&\sum_{\bk \sigma} (\enk-h\sigma) \cdks \cks
    -\sum_{\bk}\left(\Delta_{\bm k}\cdku\cdmkd
    + {\rm H. c.}\right)
\nonumber \\
    &&-J \sum_{\bm k} \left[ m_{\bm Q} \cdkd \ckQu + {\rm H. c.} \right]+\frac{|\Delta_0|^2}{V}+ J |m_{\bm Q}|^2\,,
    \nonumber \\
\end{eqnarray}
with a tight-binding dispersion $\varepsilon_{\bm k}=2t(\cos k_x+\cos k_y)-\mu$. We fix
$\mu/t=0.749$
which sets the wave vectors connecting the nodal points to
${\bm Q} = (\pm 0.88\pi, \pm 0.88 \pi)$. The order parameter $\Delta_{\bm k}= \Delta_0 \eta_{\bm k}$  with $\eta_{\bm k}=\cos k_x - \cos k_y$
and
$\Delta_0=V\sum_{\bm k}\eta_{\bm k} \langle c_{-\bm k \downarrow}c_{\bm k  \uparrow} \rangle$.
Again we
first consider a single-$\bm Q$ XY magnetic ordering.
We diagonalize ${\cal H}$ and minimize the free energy to obtain the phase diagram as a function $h$ and $T$ for given interactions $J$ and $V$. The normal state becomes unstable towards magnetic ordering at
$J_c=3.6 t$.
 Consequently, we choose
 $J=3.5t$ to describe a system with strong AFM tendencies.
The phase diagram for $V=3.0t$ in Fig.~\ref{fig:PD} shows the coexistence of AFM and SC (Q-phase) below $T_{\bm Q}<T^\star$ at high fields as well as the first order N to SC transition at $0<T_c (h)< T^\star$. The latter appears here due to purely Zeeman coupling to the field, dominant at high fields for Pauli-limited CeCoIn$_5$, but is known to persist in the presence of moderate orbital effects~\cite{HAdachi:2003}. The phase diagram is qualitatively the same for $0.9 J_c \lesssim J < J_c$ and is 
robust against changes of $V$ and the inclusion of the next-nearest neighbor hopping $t'$. This strong coupling result confirms the instability analytically found in the WCL [see Eq.(\ref{eq:TQ})], and shows that the phase diagram is ubiquitous for systems such as CeCoIn$_5$.
%

The above analysis
demonstrates the instability of a Pauli limited $d$-wave superconductor towards incommensurate AFM. We now determine the most stable magnetic order. If the nodes of the order parameter are located at $(\pm 1/\sqrt 2, \pm 1/\sqrt 2)k_F$, there are four distinct wave-vectors, ${\bm Q}_i = (\pm \sqrt{2}, \pm \sqrt{2}) k_F$ that connect the nodal quasiparticle pockets.
%
The most general magnetic structure in the SDW state is $m({\bm r})\equiv m_x(\bm r)+ im_y(\bm r)= \sum_{ j=1,2 } \left[ m_{{\bm Q}_j} e^{i {\bm Q}_j \cdot {\bm r}} + m_{{\bm {\bar Q}}_j} e^{i {\bm {\bar Q}}_j \cdot {\bm r}} \right]$ (recall the spin $z$-axis is along $\bm H$). As CeCoIn$_5$ has an
easy $c$-axis (taken as spin $x$-axis), the state with $m_y=0$ is energetically more favorable.
Now consider each single-${\bm Q}$ term $c^\dagger_{{\bm k}+{\bm Q} \uparrow}\ckd$.
Recall that we pair spin-up Bogoliubov quasiparticles and that the electron creation (annihilation) operator at ${\bm k}$ is a linear combination of the Bogoliubov creation (annihilation) operator
with the same momentum and spin, and an annihilation (creation) operator with the opposite spin and momentum. As Fig.~\ref{fig:pairs} shows,
for $\bm k$, for example, in the pocket centered around $-{\bm k}_{n1}$, 
Eq.(\ref{eq:H_AFM}) produces a logarithmically divergent pairing interaction {\it only} for ${\bm Q}={\bm Q}_1 = 2 {\bm k}_{n1}$. In contrast, if
$\bm k$  is in the vicinity of ${\bm k}_{n1}$, only ${\bm Q}={\bm {\bar Q}}_1 = -2 {\bm k}_{n1}$
is relevant for the AFM instability. Similar analysis applies to the other pair of pockets. Since ${\bm Q}_i$ are far
from $(\pi,\pi)$, {\it in the WCL} the equations for the four Fourier components ${\bm m}_{{\bm Q}_i}$ decouple, look identical, and hence $|m_{ {\bm Q}_i}|=m_0$ for all ${\bm Q}_i$.
The condition $m_y=0$ immediately yields a double-${\bm Q}$ {\em collinear} structure,
${\bm m}({\bm r})= {\bm m}_0 [\cos{({\bm Q}_1\cdot {\bm r}+\varphi_1)}+\cos{({\bm Q}_2\cdot {\bm r}+\varphi_2)}] $,
with $\bm m\| \widehat{\bf c}$ and simultaneous modulation of SDW along the $[110]$ and $[1\bar{1}0]$ directions.
%
%
Note that the stabilization of this
structure is simply due to the counting of the minimal number of independent Fourier components of the order parameter that are needed to fully gap the Fermi pockets of Bogoliubov quasiparticles. The same counting leads to the 
single-${\bm Q}$ collinear phase of Cr ~\cite{EFawcett:1988}.

Beyond WCL, as states away from the Fermi surface participate in AFM, generally single-${\bm Q}$ structures become advantageous~\cite{Fazekas}. In CeCoIn$_5$ elastic neutron scattering~\cite{MKenzelmann:2008} finds a single-${\bm Q}$ modulation with $\bm Q\perp \bm H$ for ${\bm H}$ along a nodal direction $[1\bar{1}0]$, and NMR lineshape at $H>10.2$T is consistent with a single-${\bm Q}$ structure~\cite{BLYoung:2007,GKoutroulakis:2010}. At these high fields the large size of the Bogoliubov pockets leads to an intermediate or strong coupling regime.
The first question is whether at lower fields, where pockets are small, our proposed double-${\bm Q}$ structure is realized. NMR measurements of Ref.~\cite{GKoutroulakis:2010} show an anomalous broadening of the line shape for 9.2T$\lesssim H \lesssim$10.2T and $T=70$mK. Moreover, 9.2T is precisely the field at which the amplitude of the ordered moment of the single-${\bm Q}$ structure extrapolates to zero, suggesting that another form of magnetic ordering persists in the 9.2T$\lesssim H \lesssim$10.2T region. The double-${\bm Q}$ structure is a natural candidate for
this intermediate phase because it produces a broad NMR line shape that is consistent with the experimental observation.
Detailed neutron and NMR experiments will be able to test this hypothesis. Thermodynamic signatures of the transition from double to single-${\bm Q}$ phase are expected to be weak.

For $\bm H\|$ nodes, the orbital coupling suppresses the SDW modulation with $\bm Q\perp \bm H$ at low fields.
Within the semiclassical approximation the Doppler shift, $\delta E\approx \bm v_F\cdot \bm p_s$, where $\bm p_s\perp \bm H$ is the momentum of the Cooper pairs, has opposite sign for the two pockets connected by $\bm Q$, as $\bm v_{F1}=-\bm v_{F2}$. From Eq.~(\ref{eq:H0}) this expands one of the pockets and shrinks its counterpart, breaking the nesting condition. For some Fermi surface geometries an additional field-dependent modulation may still allow SDW formation, similar to
pnictides~\cite{VCvetkovic:2009,ABVorontsov:2010} and exciton formation~\cite{LPGorkov:1972}. Generically however, AFM is favored when the typical Doppler shift $\delta E_D\sim \Delta_0\sqrt{H/H_{c2}^{orb}}\geq \mu_B H$, where $H_{c2}^{orb}$ is the orbital critical field. In CeCoIn$_5$ this gives fields $H\geq H_P^2/H_{c2}^{orb}$, where $H_P$ is the Pauli limiting field, above 20-40\% of the observed $H_{c2}(T=0)$. The semiclassical approach is inadequate at high fields~\cite{ABVorontsov:2006} where a more detailed analysis is required to determine whether the $\bm Q\perp \bm H$ state is stable. In weak coupling at low fields, SDW with $\bm Q\| \bm H$ may exist since $\delta E$
vanishes for the quasiparticle pockets around the nodes aligned with the field~\cite{IVekhter:1999}, but  $T_Q$ is exponentially small.

Anisotropic spin-spin interactions induced by the spin-orbit coupling may also favor ordering with $\bm Q\perp {\bm H}$, but depend on specific models. An intriguing possibility is that this choice is due to putative FFLO state. In the presence of vortices the FFLO modulation is along the field at a wave vector determined by the field magnitude, and interferes destructively with the AFM at $\bm Q\| \bm H$ whose wave vector is fixed by the position of the SC nodes. Hence $\bm Q\perp {\bm H}$ is stabilized. Since generically the $T_{\bm Q}(H)$ that we find differs from the onset of the FFLO modulation, one expects two transitions from the superconducting state, first to a double-$\bm Q$ AFM, followed by the FFLO+single-${\bm Q}$ state. Further experiments will distinguish between these scenarios.

To summarize, our theoretical framework describes a generic instability of Pauli-limited unconventional superconductors under Zeeman effect.
This instability leads to incommensurate AFM  at the wave-vectors connecting opposite nodal points. Our theory describes the existing experiments on CeCoIn$_5$ and suggests studies to identify the double-${
\bm Q}$ collinear phase that we find to be favorable in the weak coupling regime, and which may exist in the intermediate field range.

We are grateful to K.~Aoyama, S. Brown, J.~Flouquet, K.~Machida, Y.~Matsuda,  R. Movshovich, T.~Sakakibara and Y.~Yanase for discussions.
Work at LANL was performed under the auspices of the
U.S.\ DOE contract No.~DE-AC52-06NA25396 through the LDRD program.
I. V. acknowledges support from DOE Grant DE-FG02-08ER46492 and 
the
hospitality of ISSP (University of Tokyo) and Institut N{\'e}el/Universit{\'e} Joseph Fourier (Grenoble), where part of this work was done.


\end{document}